\documentclass[amsmath,amssymb,aps,pr,reprint]{revtex4-2}

\usepackage{graphicx}
\usepackage{booktabs, array, mathptmx, float, tabularx, booktabs, lipsum, amsmath,multirow}
\usepackage{siunitx, xcolor}
\usepackage[version=4]{mhchem}
\usepackage[colorlinks,linkcolor=red,anchorcolor=blue,citecolor=red,CJKbookmarks=True]{hyperref}
\usepackage{dcolumn}
\usepackage{bm}
\usepackage{changes}

\begin{document}

\title{Update of hadronic decays of $J/\psi$ and $\psi(2S)$ though virtual photons}

\author{Libo Liao}
 \email[ ]{liaolibo@ihep.ac.cn}
\affiliation{Guangxi Key Laboratory of Machine Vision and Intelligent Control, Wuzhou University, Wuzhou, Guangxi, 543002, China}

\author{Shudong Wang}
 \altaffiliation[Also at ]{University of Chinese Academy of Sciences}
 \author{Gang Li}
 \email[Corresponding Author: ]{li.gang@ihep.ac.cn}
\affiliation{
Institute of High Energy Physics, Chinese Academy of Sciences, Shijingshan District, Beijing, 100049, China 
}

\author{Zhaoling Zhang}
\author{Weimin Song}%
\affiliation{%
College of Physics, Jilin University, Changchun, Jilin, 130012, China
}%

\date{\today}

\begin{abstract}
The hadronic decay branching ratios of $J/\psi$ and $\psi(2S)$ through virtual photons $B(J/\psi, \psi(2S) \rightarrow \gamma^*\rightarrow \text{hadrons})$ are updated by using the latest published measurements of the $R$ value and the branching ratios of $J/\psi, \psi(2S) \rightarrow l^+l^-$. The precision in $B(J/\psi\rightarrow \gamma^*\rightarrow \text{hadrons})$ and $B(\psi(2S)\rightarrow \gamma^*\rightarrow \text{hadrons})$ improve by factors of 4 and 3, respectively. 
\end{abstract}

\maketitle

The $J/\psi(1^3S_1)$ and $\psi(2^3S_1)$,  $J^{PC}(1^{--})$ states of charmonium have extremely small width. The reason is usually referred to by the phenomenological Okubo-Zweig-Iizuka rule,  which states that processes in which the initial quark pairs cannot appear as part of the final state particles are suppressed. Therefore, the $J/\psi$ only decays through virtual photons (vacuum polarization), three gluons, two gluons plus a single photon, or into $\eta_c$ via M1 transition and three photons. The determination of the partial widths could play an important role in understanding the nature of $J/\psi$ decays. The hadronic final state through virtual photons and gluons is indistinguishable. Fortunately, the branching ratios $B(J/\psi \rightarrow \gamma^*\rightarrow \text{hadrons})$ can be calculated using the relation from Ref.~\cite{Jpsi_decays}

\begin{equation}
    \begin{split}
        B(J/\psi \rightarrow \gamma^*\rightarrow \text{hadrons}) &= \frac{\sigma_h}{\sigma_l}\times B(J/\psi \rightarrow l^+l^-)\\
        & \equiv R \times B(J/\psi \rightarrow l^+l^-)~~~~~(l=e,\mu),
    \end{split}
    \label{equ:relation}
\end{equation}
where assumes that the ratios of the cross sections for virtual photons decaying to hadrons and leptons, denoted as $\sigma_h$ and $\sigma_l$ respectively, are the same for both electron-positron and $c\bar{c}$ annihilations. This relationship also holds for the decay of $\psi(2S)$.

The best published determinations on $B(J/\psi, \psi(2S) \rightarrow \gamma^*\rightarrow \text{hadrons})$ are from 2004 and were determined by Seth~\cite{Seth:2004qc}:
\begin{equation}
    B(J/\psi\rightarrow \gamma^*\rightarrow \text{hadrons}) = (13.5\pm0.3)\%,
\end{equation}
\begin{equation}
    B(\psi(2S) \rightarrow \gamma^*\rightarrow \text{hadrons}) = (1.66\pm0.10)\%.
\end{equation}
The $R$ value used in this result was $2.28\pm0.04$. 

Recently, more precise results for the $R$ value and $B(J/\psi, \psi(2S) \rightarrow l^+l^-)$ became available.
The KEDR experiment measured the $R$ value between 1.84 and 3.05 GeV~\cite{Anashin:2016hmv} and between 1.84 and 3.72 GeV~\cite{Anashin:2015woa,KEDR:2018hhr}. 
The BESII measured the $R$ value at energy points of 3.650, 3.6648 and 3.773 GeV~\cite{BESII-2006-1}, between 3.650 and  3.872 GeV~\cite{BESII-2006-2}, and at 2.60, 3.07 and 3.65 GeV~\cite{BESII-2009}. 
The newest measurements by BESIII range from 2.2324 to 3.6710 GeV~\cite{BESIII:2021wib}. 

In the studies of $B(J/\psi, \psi(2S) \rightarrow l^+l^-)$, the most recent and the most precise results were published by CLEO\underline{~}$c$ and BESIII, which reached relative precision of 0.6\% and 2.1\%, respectively. Reasons for these improvements are better detectors, larger data samples, and carefully studied systematic uncertainties. 
By combining all available results~\cite{PDG-2022}, the precisions on $B(J/\psi\rightarrow l^+l^-)$ and $B(\psi(2S)\rightarrow l^+l^-)$ improved by factors of 4 and 2.5, respectively, as shown in Table~\ref{tab:llbr2022}.

\begin{table*}[hbpt]
\caption{Summary of the branching ratios $B(J/\psi, \psi(2S) \rightarrow l^+l^-)$ and their averages in PDG 2002~\cite{PDG-2002} and 2022~\cite{PDG-2022}. In the 3rd and 5th columns, the branching ratios $B(J/\psi\rightarrow l^+l^-)$ and $\psi(2S) \rightarrow l^+l^-)$ were assuming lepton universality and neglecting the mass difference between electrons and muons. The relative uncertainty of the branching ratios is also provided.}
\label{tab:llbr2022}
\begin{ruledtabular}
    \begin{tabular}{ccccccc}
             & PDG 2002 (\%)& Average (\%) & Relative uncertainty (\%) & PDG 2022 (\%) & Average (\%)& Relative uncertainty (\%)\\ \hline
        $B(J/\psi\rightarrow e^+e^-)$ & (5.93$\pm$0.10) & \multirow{2}{*}{(5.90$\pm$0.09)}&  \multirow{2}{*}{1.6} & (5.971$\pm$0.032) & \multirow{2}{*}{(5.967$\pm$0.023)} & \multirow{2}{*}{0.4}\\ 
        $B(J/\psi\rightarrow \mu^+\mu^-)$ & (5.88$\pm$0.10) &  & &(5.961$\pm$0.033) & & \\ 
        \hline
        $B(\psi(2S) \rightarrow e^+e^-)$ & (0.73$\pm$0.04) & \multirow{2}{*}{(0.73$\pm$0.04)}&\multirow{2}{*}{5.5}& (0.793$\pm$0.017)& \multirow{2}{*}{(0.794$\pm$0.017)}& \multirow{2}{*}{2.2}\\ 
        $B(\psi(2S) \rightarrow \mu^+\mu^-)$ & (0.70$\pm$0.09) & & & (0.80$\pm$0.06) & & \\ 
    \end{tabular}
\end{ruledtabular}
\end{table*}

\begin{figure*}
    \includegraphics[width=.8\textwidth]{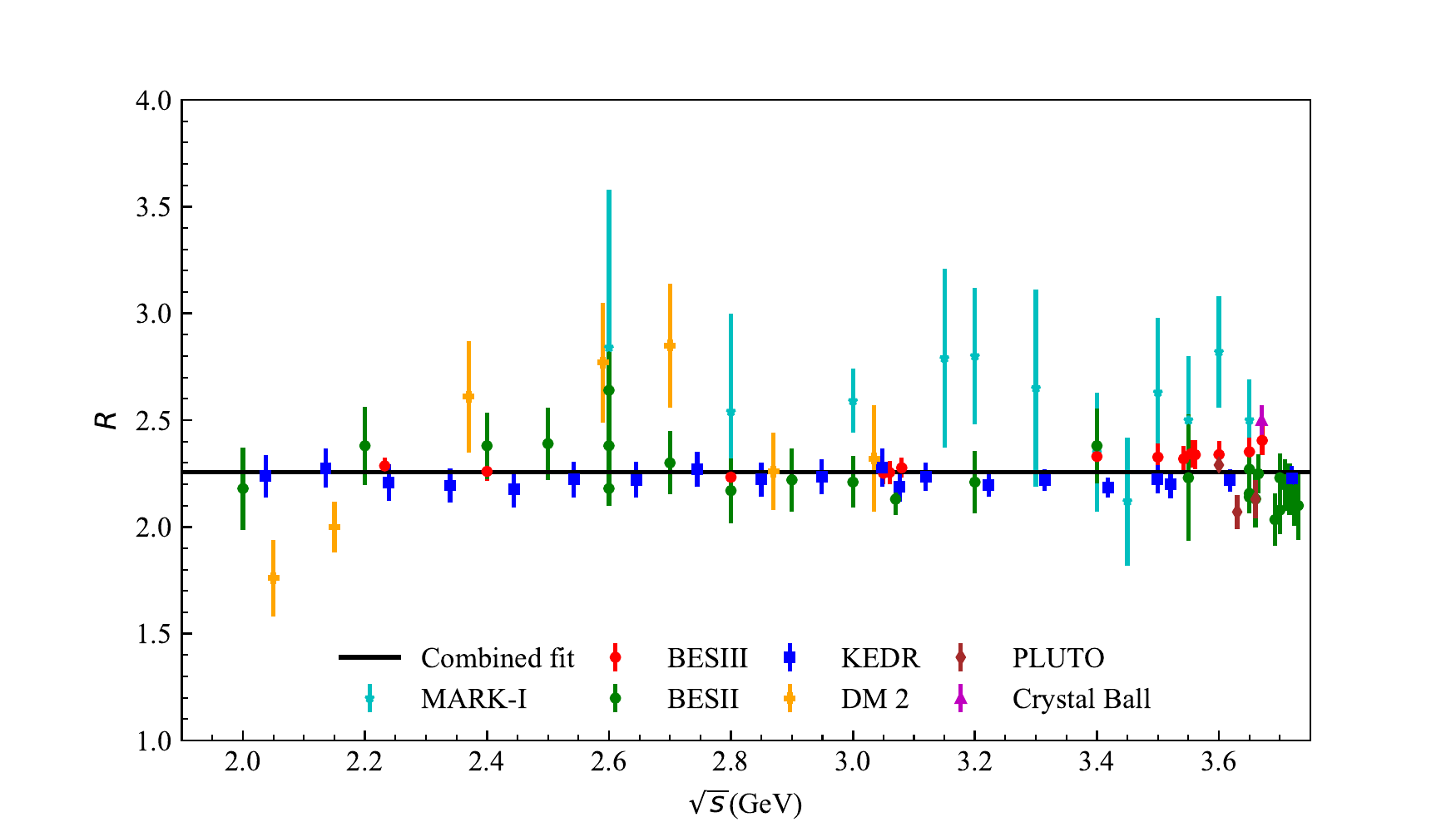}
    \caption{Summary of $R$ measurements in the c.m. energy region from 2.0 to 3.73 GeV. Red dots are results from BESIII~\cite{BESIII:2021wib}. The $R$ values of the BESII experiment~\cite{BESII-1999,BESII-2001,BESII-2006-1,BESII-2006-2,BESII-2009} are indicated as green dots. Blue rectangles represent the results of KEDR detector~\cite{Anashin:2015woa,Anashin:2016hmv,KEDR:2018hhr}. MARK-1~\cite{MARK-1-1981} and DM 2~\cite{DM2-1979} results are shown as cyan stars and orange crosses, respectively. The values from PLUTO~\cite{PLUTO-1981} are denoted by brown diamonds. Magenta triangles stand for the results of the Crystal Ball Collaboration~\cite{CrystalBall-1986}.}
    \label{fig:globalfit}
\end{figure*}

The $R$ values between 2.0 and 3.73 GeV are used to approximate the ones at $J/\psi$ and $\psi(2S)$ peaks, as shown in Fig.~\ref{fig:globalfit}. The results of the $R$ scans used in this study are from the MARK-I, the DM 2, the BESII, the KEDR, the BESIII, etc. It can be seen that the precision of $R$ values has improved significantly in the last decades, benefiting from better detectors, increasing data samples, and precise knowledge of initial state radiation correction and systematic uncertainties. 

Various combinations of the existing measurements of the $R$ value are tested to determine the $R$. The results are summarized in Table.~\ref{tab:Rscans}. It can be seen clearly that the BESII, the KEDR, and the BESIII can achieve comparable precision, and all three are better than the one used in Ref.~\cite{Seth:2004qc}. 
It should be noted that the BESII published new measurements of the $R$ value in 2006~\cite{BESII-2006-1,BESII-2006-2} and 2009~\cite{BESII-2009} after the publication of Ref~\cite{Seth:2004qc}. 
The precision reaches 0.4\% if all the results are combined, which is going to be used in the following calculations.

\begin{table}[hbpt]
\centering
\caption{$R$ value scan results from several experiments in the center of mass energy range between 2.0 to 3.73 GeV. The combined fit result is used to approximate the $R$ values underneath the $J/\psi$ and $\psi(2S)$ peaks. Only experiments on more than five energy points are included.}
\label{tab:Rscans}
    \begin{ruledtabular}
    \begin{tabular}{ccc}
         &  $R$ & Relative uncertainty\\
    \hline
        MARK-I & $2.56\pm0.08$    & 3.1\%\\
        DM 2 & $2.19\pm0.07$      & 3.3\%\\
        BESII & $2.20\pm0.02$     & 1.0\%\\
        KEDR & $2.22\pm0.02$      & 0.7\%\\
        BESIII & $2.30\pm0.01$    & 0.6\%\\
    \hline
        Combined fit & $2.26\pm0.01$ & 0.4\%\\
    \hline
     Ref.~\cite{Seth:2004qc} & $2.28\pm0.04$ & 1.8\%\\
    \end{tabular}
    \end{ruledtabular}
\end{table}

The branching ratios $B(J/\psi, \psi(2S) \rightarrow \gamma^*\rightarrow \text{hadrons})$ are calculated using equation~(\ref{equ:relation}) and the combined fit result on $R$ from Table~\ref{tab:Rscans}. The results in  
\begin{equation*}
        B(J/\psi\rightarrow \gamma^*\rightarrow \text{hadrons})  = (13.46\pm 0.07)\%~,
\end{equation*}
and
\begin{equation*}
        B(\psi(2S) \rightarrow \gamma^*\rightarrow \text{hadrons})    = (1.79\pm 0.04)\%~,
\end{equation*}
are compared to the previous results in Table~\ref{tab:branchratio}. Benefiting from the improved precision of the $R$ values and $B(J/\psi, \psi(2S) \rightarrow l^+l^-)$, the relative uncertainties are reduced significantly. The $B(J/\psi\rightarrow \gamma^*\rightarrow \text{hadrons})$ and $B(\psi(2S) \rightarrow \gamma^*\rightarrow \text{hadrons})$ are improved by about 4 times and 3 times, respectively.  But it should be noted that the correlations among R values within a single experiment are neglected, as well as the correlation between the branching ratios for $B(J/\psi \to e^+e^-)$ and $B(J/\psi \to \mu^+\mu^-)$.  

\begin{table*}[ht]
\caption{Comparison of $B(J/\psi, \psi(2S) \rightarrow \gamma^*\rightarrow \text{hadrons})$ in Ref~\cite{Seth:2004qc} and those in this study. }
\label{tab:branchratio}
\begin{ruledtabular}
    \begin{tabular}{cccc}
        Channel& Ref.~\cite{Seth:2004qc} & This study & Relative uncertainty\\
         \hline
        $B(J/\psi\rightarrow \gamma^*\rightarrow h)$ &(13.5$\pm$0.3)\%& (13.46$\pm$0.07)\% &2.2\%$\rightarrow$0.5\%\\
        $B(\psi(2S) \rightarrow \gamma^*\rightarrow h)$ &(1.66$\pm$0.10)\%&(1.79$\pm$0.04)\% &6.1\%$\rightarrow$2.1\% \\
    \end{tabular}
\end{ruledtabular}
\end{table*}

The update on $B(J/\psi, \psi(2S) \rightarrow \gamma^*\rightarrow \text{hadrons})$ will help to constrain branching ratios $B(J/\psi,\psi(2S)\rightarrow \gamma gg, ggg)$ and enhance our understanding of the two charmonia. Taking the $J/\psi$ as an example, two important aspects arise. Firstly, $B(J/\psi \rightarrow \gamma^\star \rightarrow \text{hadrons})$ can be utilized to test the theoretical calculation presented in equation (8) of Ref.~\cite{Jpsi_decays}, including its higher-order corrections. Secondly, it enables the calculation of branching ratios such as $B(J/\psi\rightarrow\gamma gg, ggg)$ and the relative fractions of $B(ggg)$, $B_{elm}(\text{had})$, $B(\gamma gg)$, and $B(\gamma\eta_c)$, in conjunction with $B(J/\psi\rightarrow\gamma\eta_c)$, $B(J/\psi\rightarrow l^+l^-)$. Notably, the precision of the ratio $\Gamma(\gamma gg)/\Gamma(ggg)$ introduces the dominant uncertainty, emphasizing the need for future updates to improve its precision.

\nocite{*}

\bibliography{sample}
\end{document}